\documentstyle[twocolumn,prb,aps,epsfig]{revtex}

\begin{document}
\draft
\title{Quantum fluctuations in the cohesive force of metallic  
nanowires} 
\author{C. H\"oppler\thanks{Present address: Institut f\"ur Angewandte 
Physik, Eber\-hard-Karls-Universit\"at T\"ubingen, Auf der Morgenstelle
10, D-72076 T\"ubingen, Germany.} and W. Zwerger}
\address{Sektion Physik, Ludwig-Maximilians-Universit\"at M\"unchen,
Theresienstra{\ss}e 37, D-80333 M\"unchen, Germany}
\date{\today}
\maketitle

\begin{abstract}
  Based on the recent free electron model for cohesion in narrow
  metallic constrictions by Stafford et al., we calculate the quantum
  fluctuations in the cohesive force versus elongation. The
  fluctuations are dominated by states near the Fermi energy, thus
  explaining their apparently universal magnitude of order
  $\varepsilon_F/\lambda_F$. We present numerical results for the
  force fluctuations in a simple geometry and show that they are well
  described by the contributions of a few classical periodic orbits in
  the Balian-Bloch trace formula for the density of states of
  transverse motion.
\end{abstract}

The mechanical properties of atomic size constrictions between two
reservoirs of standard metals, for example Au, have been the subject
of intensive research in the past few years.\cite{agrait,krans,olesen}
In particular, simultaneous measurements of the cohesive force and the
electrical conductance \cite{rubio,stalder} have shown a striking
correlation between the mechanical and transport properties. In a
regime where the cohesive force is linear in the elongation (`elastic
stage'), the electrical conductance exhibits plateaus similar to those
found in two dimensional quantum point contacts in semiconductor
heterostructures.\cite{wees,wharam} By contrast, both the conductance
and the elastic force rapidly decrease in the narrow regime between
two conductance plateaus (`yielding stage'). These observations were
originally explained \cite{rubio} in terms of abrupt atomic
rearrangements which appear with increasing elongation and indeed
classical molecular dynamics simulations
\cite{landmann,todorov,sorensen} seem to support this point of view.
On the other hand, the apparent similarity between the conductance
plateaus found here and in semiconductor quantum point contacts
together with the strong correlation between conductance and force
suggests that elastic and yielding stages in the cohesive force may
appear even for a smooth constriction geometry as a result of changing
the number of discrete transverse modes for the electrons.\cite{note1}

This suggestion has recently been developed by Stafford, Baeriswyl,
and B\"urki.\cite{stafford}  It is based on viewing the transverse
eigenstates of the conduction electrons as delocalised chemical bonds
which provide both conduction {\em and} cohesion. Remarkably, a
corresponding free electron model qualitatively accounts both for the
average cohesive force and for the abrupt steps in the force of order
$\varepsilon_F/\lambda_F$ which are caused by successively cutting off
the discrete electronic modes in the constriction.  Now it is
evident that any purely electronic model is an idealization of the
true experimental situation.  For instance, possible atomic
rearrangements which reveal themselves through a hysteretic behavior
of the force \cite{rubio} are neglected. A realistic electronic
description moreover has to account for the strong deviations from a
naive conductance quantization picture arising from the rather small
value of the Fermi wavelength which makes the electronic motion
sensitive to defects even on an atomic scale.\cite{scheer}
Nevertheless, the simple free electron model of Stafford et al. not
only provides an intuitive explanation of the observed behavior but
also gives a correct order of magnitude estimate of the relevant
forces, in the same way perhaps as the measured bulk modulus of many
metals is roughly determined by the ground state Fermi gas pressure of
the conduction electrons.\cite{ashcroft}

In our present work we discuss the fluctuations in the cohesive force
which arise from the discreteness of the electronic eigenstates for
motion in the direction transverse to the elongation.\cite{blom} As
shown by Stafford et al., the electronic cohesive force $F$ of a
constriction of length $L$ is obtained from the $T=0$ free electron
grand canonical potential
\begin{equation}\label{eq:gkPot}
\Omega=-\frac{8}{3}\frac{\varepsilon_F}{\lambda_F}\,\int_0^L{\rm
d}z\int_0^{\varepsilon_F}{\rm
d}E\,\rho_{\perp}(E,z)\,\bigl(1-E/\varepsilon_F\bigr)^{3/2}
\end{equation}
simply via $F=-\partial\Omega/\partial L$. Here $\rho_{\perp}(E,z)$ is
the differential density of states (DOS) for the transverse motion of
the electrons at a given cross section $(z=\mbox{const.})$. Within a
semiclassical description, which is valid as long as the Fermi
wavelength $\lambda_F$ is much smaller than the constriction width $R$,
the DOS may be split into an average $\bar\rho_{\perp}(E,z)$ and a
fluctuating contribution $\delta\rho_{\perp}(E,z)$. The latter
vanishes if averaged over an energy range much larger than the spacing
$\Delta(z)$ between successive transverse eigenstates. Assuming the
deformation occurs at constant total volume $V$, the average cohesive
force $\bar F$ which is associated with the average DOS has only two
contributions for $\lambda_F\ll R$: \cite{stafford,hoeppler}
\begin{equation}\label{eq:averageF}
\bar F=-\frac{\varepsilon_F}{\lambda_F}\left[\frac{k_F}{8}
\left.\frac{\partial S}{\partial L}\right|_V\, -\frac{4}{9}\right].
\end{equation}
Here the dominant contribution is associated with the (electronic
contribution to the) surface tension, giving a cohesive force
proportional to the change in surface area $S$ with elongation $L$. In
addition there is a universal contribution
$\frac{4}{9}\,\frac{\varepsilon_F}{\lambda_F}$ to the average cohesive
force which is completely independent of the geometry.\cite{hoeppler}
It arises from the topological term in the Weyl expansion for the
integrated DOS \cite{kac} and leads to a weakening of the cohesion
compared to the macroscopic surface tension contribution. As is
evident from equation (\ref{eq:averageF}), the average cohesive force
does not depend on the detailed geometry of the nanosize constriction
which will presumably vary significantly between different
realizations of an experiment.  Regarding the force fluctuations
$\delta F$ which arise from the fluctuating part of the DOS, it was
found in the numerical calculations \cite{stafford} that the precise
form of $\delta F$ is specific to the shape of the cross section but
does not depend on the detailed form of the constriction radius $R(z)$
versus $z$. Remarkably, the rms amplitude of these fluctuations turned
out to have a universal magnitude
\begin{equation}
  \label{eq:rms}
  {\rm rms}(\delta F)\approx 0.3\frac{\varepsilon_F}{\lambda_F}
\end{equation}
independent of geometry. In order to understand these observations, we
note first that $\delta F$ is determined by the fluctuations in the
DOS of {\em transverse} motion. Following recent work
\cite{blom,ruitenbeek,kassubek} we may therefore assume a simple
cylindrical form $R(z)=\mbox{const.}$ for the constriction with a
radius $R$ which scales like $L^{-1/2}$ in order to fulfill the
constraint of constant volume. Since the discrete eigenenergies
$\varepsilon_{\nu}$ for transverse motion all scale like $R^{-2}$, we
have $\partial\varepsilon_{\nu}/\partial L=\varepsilon_{\nu}/ L$. It
is then easy to show from (\ref{eq:gkPot}) that in this simple
geometry the fluctuations of the cohesive force at fixed total volume
are given by
\begin{equation}\label{eq:deltaF}
  \delta F |_V=\frac{8}{3}\frac{\varepsilon_F}{\lambda_F}\,
  \int_0^{\varepsilon_F}\!{\rm d}E\,\delta\rho_{\perp}(E)\!
  \left(1-\frac{5}{2}\frac{E}{\varepsilon_F}\right)
  \!\left(1-\frac{E}{\varepsilon_F}\right)^{1/2}.  
\end{equation}
Here $\delta\rho_{\perp}(E)$ is the fluctuating transverse DOS at the
narrowest point of the constriction, which also determines its
conductance $G$. Now, in contrast to $G$ which is a property of the
states directly {\em at} the Fermi energy, the cohesive force
obviously depends on {\em all} the states with energies between zero
and $\varepsilon_F$. The numerical calculations
\cite{stafford,kassubek}, however, which show that the force
oscillations are directly correlated with the conductance, indicate
that $\delta F$ is dominated by the states near the Fermi energy.
Indeed, this observation can be understood easily from equation
(\ref{eq:deltaF}), at least on a qualitative level.  Since the
fluctuating DOS $\delta\rho_{\perp}(E)$ is a rapidly oscillating
function which varies on a scale of the order of the mean level
spacing $\Delta\ll\varepsilon_F$, the contributions to $\delta F |_V$
in (\ref{eq:deltaF}) from energies between zero and close to
$\varepsilon_F$ cancel.  It is only in a small range of several level
spacings near the upper integration limit, where the factor
$(1-E/\varepsilon_F)^{1/2}$ changes rapidly on a scale on which the
DOS varies. Therefore only the contribution to $\delta F$ from a few
states below the Fermi energy survives in (\ref{eq:deltaF}). As a
result, $\mbox{rms}(\delta F)$ is expected to be of order
$\varepsilon_F/\lambda_F$ independent of the constriction radius $R$,
a property which has been verified numerically up to values $k_FR=200$
by Stafford et al.\cite{stafford,note2} By contrast, if all the states
from zero up to $\varepsilon_F$ were to contribute to the force
fluctuations, $\delta F$ would scale as the fluctuations in the {\em
  total} number of states $N(\varepsilon_F)$ below the Fermi energy.
Assuming Poisson statistics, appropriate for a classically integrable
transverse motion \cite{haake}, one has
$\mbox{rms}N(\varepsilon_F)=\bar N^{1/2}(\varepsilon_F)\sim k_FR$. The
force fluctuations would thus increase with the constriction radius,
in contradiction with the results of Stafford et al.\cite{stafford}

\begin{figure}[tbp]
  \epsfig{file=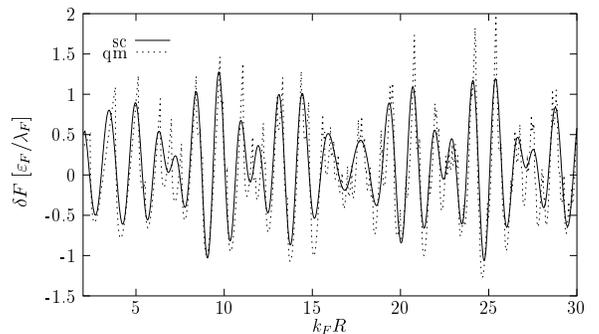, width=0.45\textwidth}
  \caption{Comparison of the fluctuating part of the cohesive force
    $\delta F$ with $\delta F^{sc}$, calculated with three periodic
    orbits ($v=2, 3, 4$, $w=1$).} 
  \label{fig1}
\end{figure}

\begin{figure}[tb]
  \setlength{\unitlength}{1cm}
  \begin{center}
    \begin{minipage}{.95\textwidth}
      \begin{minipage}{.95\textwidth}
        \epsfig{file=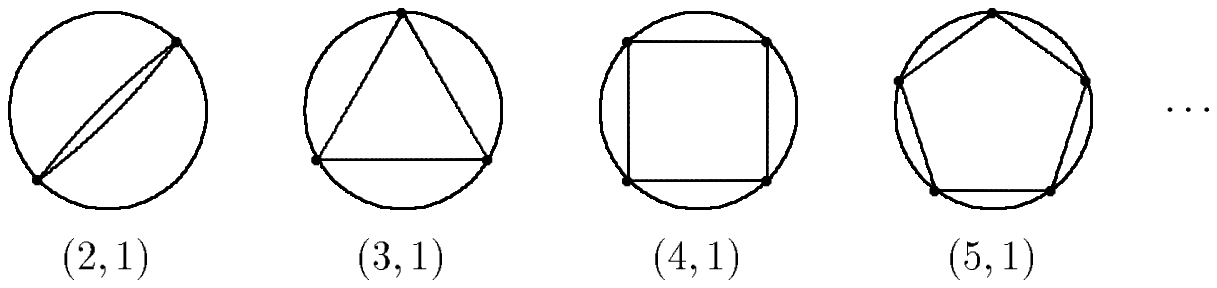, width=0.5\textwidth}
      \end{minipage}
      \begin{minipage}{.95\textwidth}
        \begin{picture}(13,2.2)
          \epsfig{file=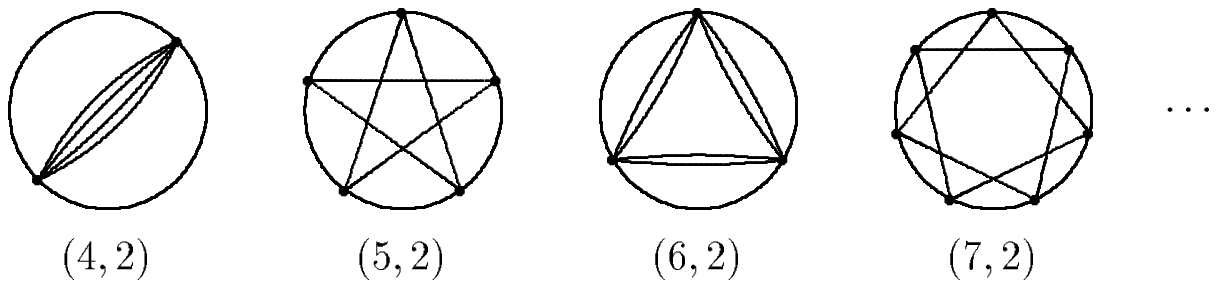, width=0.5\textwidth}
        \end{picture}
      \end{minipage}
      \begin{minipage}{0.95\textwidth}
        \begin{picture}(10,2.2)
          \epsfig{file=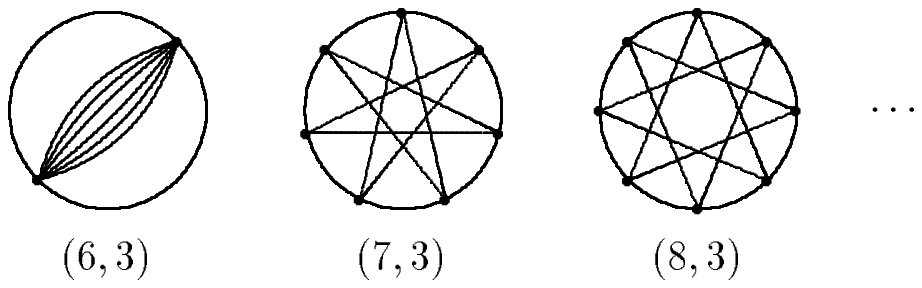, width=0.38\textwidth}
        \end{picture}
      \end{minipage}
    \end{minipage}   
    \vspace{2mm}
    \caption{Closed classical periodic orbits in a circlular billiard
      with reflecting walls. Between consecutive reflections the
      trajectory follows straight lines.  The winding number $w$ and
      the number of vertices $v$ of the particular orbits are given as
      tupel $(v,w)$ (after Balian and Bloch \cite{balian}).}
    \label{fig2}
  \end{center}
\end{figure}

For a quantitative confirmation of our arguments above, we have
calculated numerically the force fluctuations which follow from
(\ref{eq:deltaF}) for a cylindrical constriction with radius
$R(L)=R_0\sqrt{L_0/L}$. The corresponding results for $\delta F$ in
units of the fundamental force $\varepsilon_F/\lambda_F\approx 1\,
\mbox{nN}$ are shown in Fig.~\ref{fig1} for a wire which is
stretched from $k_FR=30$ down to $k_FR=2$. Clearly, the force
fluctuations are independent of $k_FR$, with an average magnitude
$\mbox{rms}(\delta F)=0.6\,\varepsilon_F/\lambda_F$ which is of the
same order as the topological contribution to the force
(\ref{eq:averageF}).\cite{note3} The apparently complicated dependence
of the force fluctuations on the elongation can in fact be simply
understood as arising from only a few classical periodic orbits of
electrons in the assumed circular cross section of the wire. Indeed,
as is well known, the fluctuations in the DOS around its average value
can generally be represented in terms of a sum over classical periodic
orbits in a semiclassical approximation. For integrable systems such a
connection between the quantum mechanical DOS and classical mechanics
was first found by Balian and Bloch.\cite{balian} Specifically, for a
circle the periodic orbits are regular polygons. They may be
characterized by their number of vertices $v$ and their winding number
$w$. Obviously we have $v\geq 2w$. If there is a common divisor $n$
between $v$ and $w$, the orbit is an $n$-fold repetition of a
primitive periodic orbit (see Fig.~\ref{fig2} for some elementary
examples). Introducing an angle $\phi_{vw}=\pi w/v$, the length of a
periodic orbit is $L_{vw}=2vR\sin\phi_{vw}$ from simple geometry. The
oscillating contribution to the DOS of a circular billiard can then be
represented as \cite{reimann}
\begin{eqnarray}
  \label{eq:deltarho}
  \delta\rho_\perp^{sc}(E)&&=\frac{2}{\Delta}(\pi
  k_ER)^{-1/2}\sum_{w=1}^\infty\sum_{v=2w}^\infty f_{vw}
  \frac{\sin^{3/2}\phi_{vw}}{\sqrt v}\nonumber\\ 
  &&\times\sin\left(k_EL_{vw}-3v\frac{\pi}{2}
  + 3\frac{\pi}{4}\right). 
\end{eqnarray}
Here $\Delta=\frac{\hbar^2}{mR^2}$ is the unit of energy and
$f_{vw}=1$ for $v=2w$ or $f_{vw}=2$ for $v>2w$ the number of different
periodic orbits through an arbitrary point within the circle. Using
the semiclassical approximation (\ref{eq:deltarho}) in our expression
(\ref{eq:deltaF}) for the force oscillations, we find that the details
of the exact numerical result for $\delta F$ are essentially explained
by including only the three simplest periodic orbits $v=2,\: 3,\: 4$,
$w=1$ in the circle (see Fig.~\ref{fig1}).  Extending the series to
all the 16 orbits with $v\leq 10$ and $w\leq 2$, the agreement between
the semiclassical and fully quantum mechanical calculation becomes
essentially exact. The fact that only a few periodic orbits are
required to describe the force oscillations is a consequence of the
integration over all energies in (\ref{eq:deltaF}), which suppresses
the DOS fluctuations on very short scales associated with longer
periodic orbits.

Finally, it is interesting to point out that the approximation of
simply adding the Weyl and trace formula contribution to the DOS,
which apparently works very well for our present problem, is not
always valid as has been shown very recently by Bhaduri et
al.\cite{bhaduri}

In conclusion, we have studied the quantum fluctuations in the
cohesive force of metallic nanowires which arise from the discreteness
of the electronic motion in the transverse direction.  It has been
shown that only a few states below the Fermi energy contribute to
these fluctuations, supporting the prediction of universal force
fluctuations $\mbox{rms}(\delta F)=\mbox{const.}\cdot
\varepsilon_F/\lambda_F$ first found by Stafford et al.\cite{stafford}
Unfortunately even in our simple geometry we have not found an
analytical derivation of this result.  Indeed, in our semiclassical
approach, if we substitute equation (\ref{eq:deltarho}) into
(\ref{eq:deltaF}) and then change the integration variable to
$E/\Delta$ it is not evident that the fluctuations are independent of
$k_FR$.  Therefore the issue of force fluctuations and in particular
their geometry dependence deserves further investigation.  The
detailed structure of the force fluctuations reflects the classical
periodic orbits in a given cross section which we have assumed to be
circular here. It would clearly be of interest to generalize our
results to the case of chaotic motion in the transverse direction
where the eigenvalues obey random matrix theory rather than Poisson
statistics.  Similarly to the situation of persistent currents in
ballistic billiard structures, we expect that the force fluctuations
will be {\em smaller} if the transverse motion is chaotic.\cite{oppen}
Experimentally, the observation of the quantum fluctuations in the
cohesive force as well as the closely related charge fluctuations of
order $e$ predicted very recently by Kassubek et al. \cite{kassubek}
would constitute a crucial test of the electronic model of cohesion in
metallic nanowires as opposed to a classical mechanical model.

We acknowledge David A. Wharam for carefully reading the manuscript.

\vspace{1cm}

\begin{references}
\bibitem{agrait} N.~Agra\"{\i}t, G.~Rubio, and S.~Vieira,
Phys. Rev. Lett. {\bf 74}, 3995 (1995).

\bibitem{krans} J.~M. Krans et al., Nature {\bf 357}, 767 (1995).

\bibitem{olesen} L.~Olesen et al., Phys. Rev. Lett. {\bf 72}, 2251
(1994).

\bibitem{rubio} G.~Rubio, N.~Agra\"{\i}t and S.~Vieira,
Phys. Rev. Lett.  {\bf 76}, 2302 (1996).

\bibitem{stalder} A.~Stalder and U.~D\"urig, Appl. Phys. Lett.  {\bf
68}, 637 (1996).

\bibitem{wees} B.~J. {van Wees} et al., Phys. Rev. Lett. {\bf 60},
848 (1988).

\bibitem{wharam} D.~A. Wharam et al., J. Phys. C {\bf 21}, L209 (1988).
  
\bibitem{landmann} U.~ Landmann et al., Phys. Rev. Lett. {\bf 77},
  1362 (1996).

\bibitem{todorov} T.~N. Todorov and A.~P. Sutton,
  Phys. Rev. Lett. {\bf 70}, 2138 (1993), Phys. Rev. B {\bf 54},
  R14234 (1996).

\bibitem{sorensen} M.~R. S\o renson, M.~Brandbyge, and K.~W. Jacobsen,
  Phys. Rev. B {\bf 54}, 3283 (1998).

\bibitem{note1} In fact, recent density functional calculations by
  H.~H\"akkinen and M.~Manninen, Europhys. Lett. {\bf 44}, 80 (1998)
  indicate that changes in the number of transverse electronic modes
  may entail atomic rearrangements.

\bibitem{stafford} C.~A. Stafford, D.~Baeriswyl and J.~B\"urki,
Phys. Rev. Lett. {\bf 79}, 2863 (1997).

\bibitem{scheer} E.~Scheer et al., Phys. Rev. Lett. {\bf 78}, 3535
  (1997) and Superlattices and Microstructures {\bf 23}, 747 (1998).

\bibitem{ashcroft} N.~W. Ashcroft and N.~D. Mermin, {\em Solid State
Physics} (Holt, Rinehart and Winston, New York, 1976), chapter 2.

\bibitem{blom} S.~Blom et al., Phys. Rev. B {\bf 57}, 8830 (1998).

\bibitem{hoeppler} C.~H\"oppler and W.~Zwerger, Phys. Rev. Lett. {\bf
80}, 1792 (1998).

\bibitem{kac} M.~Kac, Am. Math. Monthly {\bf 73}, 1 (1966), for a
  recent discussion see e.~g. G.~Guiterrez and J.~M. Yanez,
  Am. J. Phys. {\bf 65}, 739 (1997).\label{kac}
  
\bibitem{ruitenbeek} J.~M. van Ruitenbeek et al., Phys. Rev. B {\bf
    56}, 12566 (1997).

\bibitem{kassubek} F.~Kassubek, C.~A. Stafford and H.~Grabert,
  cond-mat/9809318 (1998).

\bibitem{note2} Note that their dimensionless Sharvin conductance
  $G_S=\frac{h}{2e^2}\,G$ is just equal to the average number $\bar
  N_\perp(\varepsilon_F)$ of transverse states at the narrowest point
  below the Fermi energy, which is $(k_FR/2)^2$ to leading order
  ($k_FR\gg 1$) for a circular cross section, see ref. \ref{kac}.

\bibitem{haake} See e.~g. F.~Haake, {\em Quantum Signatures of Chaos}
  (Springer, Berlin, 1991).
  
\bibitem{note3} The rms of the force fluctuations is larger in this
  model than in that of Stafford et al.---cf. equ.
  (\ref{eq:rms})---because the smooth constriction geometry they
  considered leads to a rounding off of the peaks of the force
  fluctuations.

\bibitem{balian} R. Balian and C. Bloch, Ann. Phys. {\bf 69}, 76 (1972).

\bibitem{reimann} S.~M. Reimann et al., Phys. Rev. A {\bf 53}, 39
  (1996), see also M.~Brack and R.~K. Bhaduri, {\em Semiclassical
    Physics} (Addison Wesley, Reading, Massachusetts, 1997), chapter 6. 
  
\bibitem{bhaduri} K.~H. Bhaduri et al., chao-dyn/9803015 v2 (1998).

\bibitem{oppen} F.~v. Oppen and E.~K. Riedel, Phys. Rev. B {\bf 48},
  9170 (1993).
\end{references}
\end{document}